          [2001/04/25 1.1 (PWD)]
\documentclass[a4paper]{esapub2005} 
\pdfoutput=1

\usepackage{times}
\usepackage{natbib}
\usepackage{graphicx}

\title{On the Physical Quantitative Assessment of Model-Based PolSAR Decompositions}
\author[*]{J. David Ballester-Berman}
\author[**]{Thomas L. Ainsworth}
\author[*]{Juan M. Lopez-Sanchez}
\affil[*]{Universitat d'Alacant, Spain\\
E-mail: davidb@ua.es, juanma.lopez@ua.es}
\affil[**]{U.S. Naval Research Laboratory, USA\\
E-mail: tom.ainsworth@nrl.navy.mil \vskip 0.2 cm {\small\today}}

\begin{document}


\maketitle

\begin{abstract}
The performance of model-based decomposition approaches rooted in the Freeman-Durden concept is still an active research line in PolSAR field according to the considerable attention it has deserved along the last twenty years. Yet, certainly, most of subsequent proposals have been driven by the only objective of getting a better qualitative balance among scattering mechanisms according to theoretical expectations. This idea is not a negative aspect per se, as for example has led to a more rigorous understanding of orientation effects in both urban and natural areas and hence to improved land cover classifications. However, an in-depth quantitative analysis on the output parameters is usually lacking in this topic. Indeed, the attention has been mostly paid to the power of dominant contributions, whereas the accuracy and interpretation of those parameters useful for practical applications (such as the effective dielectric constants derived from alpha and beta parameters or vegetation randomness and orientation) have been almost systematically overlooked. The questions that remain to be answered are: What is the actual role of all parameters describing the models? Can we assign them a consistent physical interpretation or are some of them acting just as fitting parameters?

The present work aims to promote the discussion on these open issues regarding the quantitative assessment of model-based PolSAR decomposition schemes. To proceed with, as a initial step we have simulated the coherency matrix according to one of the general models proposed in the literature. Then, the inversion performance has been analysed in terms of the histograms of output parameters, standard deviation and bias. Simulations were carried out for different entropy scenarios. The analysis reveals that even the backscattering powers associated with all three basic scattering mechanisms are estimated with a non-negligible error higher than 10\% for some cases. Despite these conclusions are subject to a particular model and inversion approach they suggest that a careful consideration of physically-based decompositions outcomes should be taken.
\end{abstract}

\section{Introduction and Motivation}

The contribution of the Freeman-Durden decomposition to the progress of radar techniques is undoubtedly acknowledged by the community. After more than twenty years since the publication of the work by Freeman and Durden there has been a large amount of publications derived from it which have also contributed to the development of the field \cite{arxiv:li2019}. As Freeman and Durden conveniently stated in their original paper this approach is expected to perform best when either $f_s$ or $f_d$ are close to zero, or when $\alpha$ or $\beta$ are close to -1 or 1. Consequently, some of those later publications were envisaged to overcome such limitations. As for the writing of this manuscript, the latest published contribution on the topic is by Singh \emph{et al.} \cite{ar:singh2019} which accounts for seven scattering mechanisms. 

Concerning the quantitative analysis of results, very few studies have hitherto performed an in-depth study of the incoherent model-based decomposition concept for quantitative remote sensing applications. Some noteworthy examples are the works by Jagdh{\"u}ber et al. \cite{ar:jagdhuber2015}, Huang et al. \cite{ar:huang2016}, Di Martino et al. \cite{ar:dimartino16}, and He et al. \cite{ar:he2016} focused on soil moisture inversion but, however, the general validation methodology mostly employed in PolSAR decomposition studies has been driven by the assessment of the \emph{best} balance among backscattering powers. Therefore, here we claim that the performance of model-based decomposition approaches for parameter retrieval is still an open issue in PolSAR field. Here we emphasize the necessity of analysing the estimation accuracy of the whole set of parameters involved in any \emph{new} physical model in order to unambiguously identify the unique and novel contributions of any proposal in terms of the final application.

The most common strategy employed to assess the performance of remote sensing techniques is based on the direct comparison of some of the output parameters (i.e. soil moisture, vegetation height, or any other) with a ground-truth data set collected simultaneously. On the other hand, it happens that the retrieved values of some of the parameters can only be interpreted in terms of the expected behaviour according to theoretical foundations. In addition, due to modelling issues there appear some parameters whose validation and interpretation are subject to a high ambiguity, as it is the case of vegetation orientation and randomness. These parameters are widely used for characterising the scattering from vegetated covers but it can be stated that their interpretation is merely supported on qualitative features extracted by visual inspection of the arrangement of vegetation elements in the scene. However, it seems not to be an obstacle in many publications to accept uncritically this apparent matching to describe the vegetation morphology. 

An alternative to overcome the lack of ground-truth data for validation is by means of simulated data. In \cite{ar:xie16} we took as a starting point the progress made by several previous contributions based on the original Freeman-Durden approach in order to carry out a quantitative analysis on a particular decomposition approach. In that paper we focused on the general model proposed by Chen et al. \cite{ar:chen2014} as we considered it included several previous improvements reported in the literature. It allows us to use all nine elements of the coherency matrix and considers four different types of volume scattering models (see Section \ref{s:met}).

We showed in \cite{ar:xie16} that a reasonably overall accuracy can be achieved by including several improvements throughout the inversion procedure based on a numerical optimisation. However, only some particular cases were considered and the whole range of input values and the different combinations among them were not employed for such purpose. Therefore, in \cite{pro:polinsar2019} we replicated the same procedure but with a wider range of entropy scenarios. Simultaneously,  Ainsworth et al. \cite{pro:ainsworth18} also pointed out the need of analysing the consistency of such models to better define their ranges of applicability. 

\begin{figure*}[h]
 \centering
\begin{tabular}{c}
  \includegraphics[width=16cm]{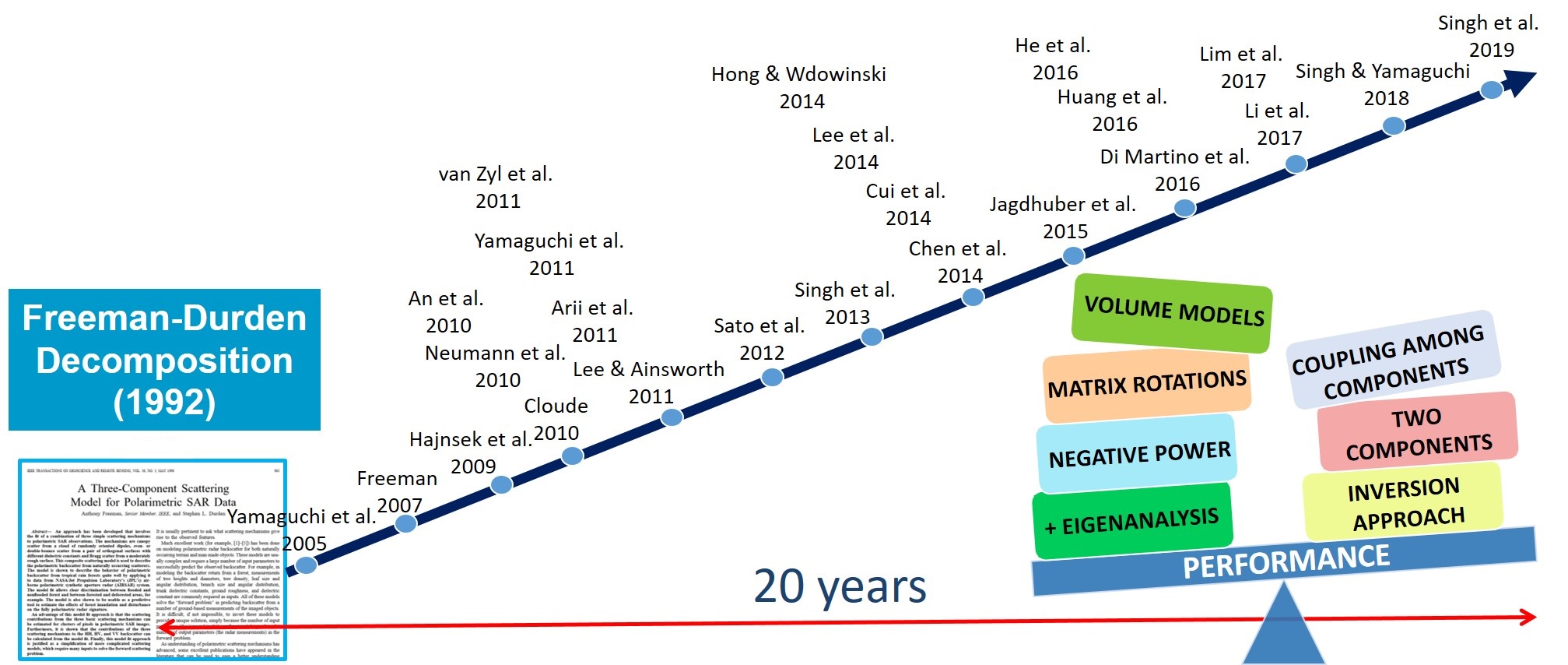}
\end{tabular}
\caption{Timeline of model-based decomposition topic since Freeman-Durden approach}
\label{f:evol}
\end{figure*}

In the present work some of the results from \cite{pro:polinsar2019} are reproduced in order to promote the discussion on the potential limits of applicability of this approach. Even though not conclusive, since they are referred to one particular model and its associated inversion method, these results allow us to speculate on the validity of the physical interpretation of some of the model parameters being employed in physically-based PolSAR approaches. It is noted, however, that after Chen's approach there appeared other new and notable contributions to the topic including different modifications and/or improvements which further contributed to the successful exploitation of PolSAR data. Figure \ref{f:evol} schematically illustrates the progress on the topic by showing some of the most relevant approaches published up to the last proposal so far \cite{ar:singh2019}. For a detailed survey on this evolution the reader is referred to the compilation in \cite{arxiv:li2019}.

Bearing in mind the previous statements, we acknowledge that the outcomes from the present work are not intended to provide any categorical conclusions on the general reliability of model-based decomposition but to suggest the need of developing parameter assessment analyses on both existing and new model-based decomposition schemes.

Section \ref{s:met} describes the methodology we have followed for the simulation study and Section \ref{s:sim} analyses some of the results that were presented in \cite{pro:polinsar2019}. Section \ref{s:disc} provides a discussion on the main points of the manuscript and on the possible strategies to progress in the field.


\section{Methodology}
\label{s:met}

\begin{figure*}[h]
 \centering
\begin{tabular}{c}
  \includegraphics[width=12cm]{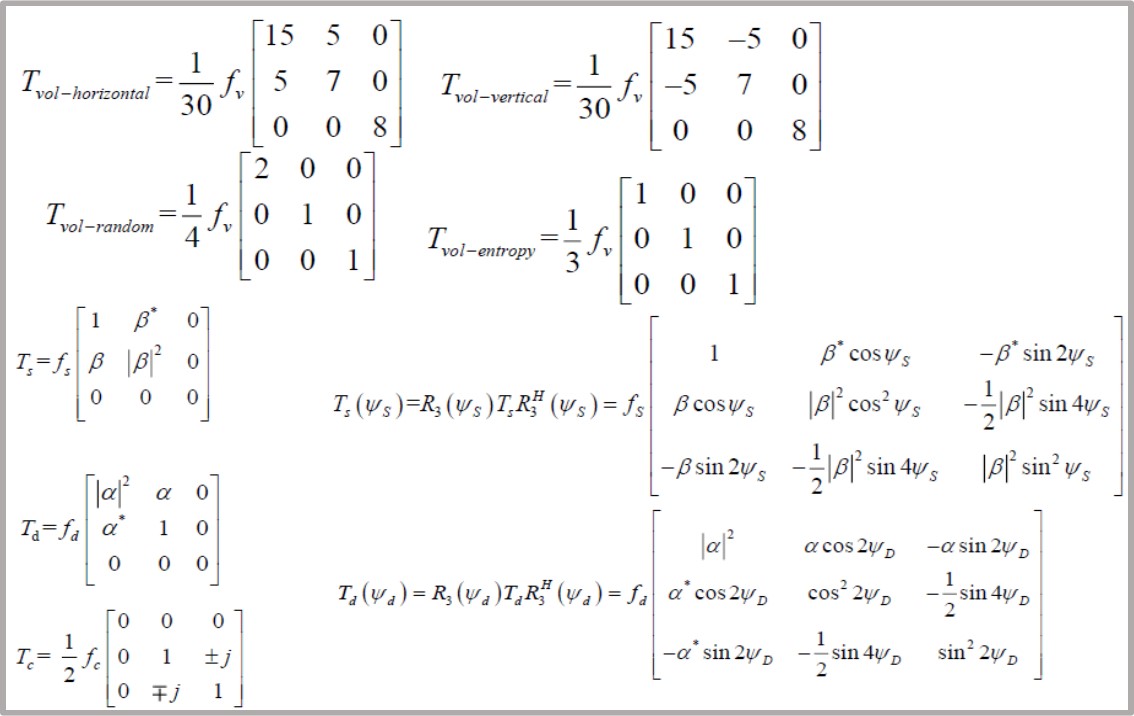}
\end{tabular}
\caption{Scattering mechanisms considered in the model-based decomposition by Chen \emph{et al.} \cite{ar:chen2014}}\label{f:chen}
\end{figure*}

The methodology employed in the present work consists on simulating the coherency matrix according to the general model proposed by Chen et al. \cite{ar:chen2014}. Three main scattering mechanisms are assumed (helix component is neglected here) where the volume is considered as a set of randomly oriented dipoles. The general expression assumed for these simulations is the following one:

\begin{equation}
T= T_v + T_d \left(\Psi_d\right)+T_s\left(\Psi_s\right) + T_{residual} \label{eq:T}
\end{equation}

where $T_v$, $T_d$, and $T_s$ correspond to the volume, double-bounce and surface scattering, respectively, and $\Psi_d$ and $\Psi_s$ represent the orientation angles for double-bounce and surface mechanisms.  Figure \ref{f:chen} displays the scattering components employed in Chen's approach.

Our hypothesis assumes that an increase of the entropy would lead to a decrease of the parameter estimation performance which certainly is an expected behaviour. But, however, the point is to ascertain those levels of entropy for which the parameter estimation becomes unreliable. This methodology is based on the following steps: 
\begin{enumerate}
\item Simulation of the coherency matrix $T$ according to Eq.(\ref{eq:T}) \cite{ar:chen2014}
\item Generation of noisy samples by applying Lee's method \cite{b:lee} from the input coherency matrix $T$.
\item Estimation of the whole set of parameters according to the inversion procedure shown in Xie et al. \cite{ar:xie16}. The MATLAB code is available in\\ \emph{https://personal.ua.es/en/davidb/soil-moisture-variations-in-deforested-tropical-areas-preliminary-results.html}
\item Computation of histograms of output parameters, standard deviation and bias with respect to the input values.
\end{enumerate}

\section{Simulations}
\label{s:sim}

Both low and high entropy scenarios were simulated according to the model and methodology described above. Note that numerical values employed for both the soil dielectric constant $\epsilon_{soil}$ and the incidence angle $\theta_0$ are given in the caption of the corresponding figure for each case.

\begin{figure*}[h]
 \centering
\begin{tabular}{c}
  \includegraphics[width=16cm]{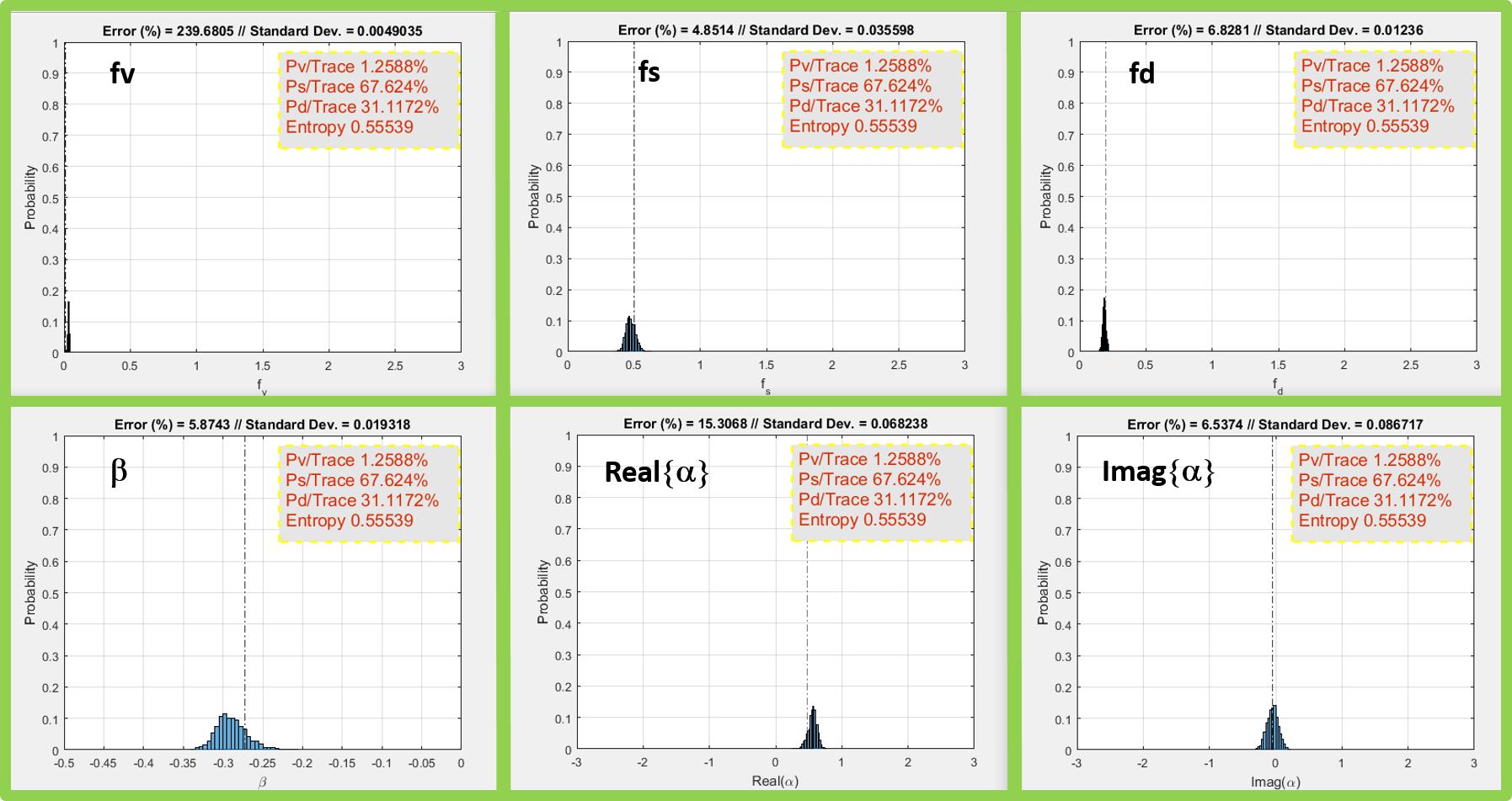}
\end{tabular}
\caption{Low entropy case. Boxes inside the plots indicate the relative power for each scattering mechanism and the entropy value. Input parameters: $\epsilon_{soil}=5$, $\theta_0=45^{\circ}$, $\Psi_d=15^{\circ}$ and $\Psi_s=10^{\circ}$}\label{f:low_entropy_1}
\end{figure*}

Firstly, low entropy cases were simulated. Figure \ref{f:low_entropy_1} shows the histograms of the estimated parameters (orientation angles $\Psi_d$ and $\Psi_s$ not shown here) for one case. At the top of each plot the mean estimation error and the standard deviation are given. This particular case assumed a negligible volume component (i.e. around 1\%) whereas surface and double-bounce scattering were about 68\% and 31\% of the total power, respectively. Only the real part of $\alpha$ and $\Psi_s$ angle (not shown here) are poorly estimated with 15.3\% and 13\% errors. The relative backscattering powers (i.e. $P_s/SPAN$ and $P_d/SPAN$) are accurately retrieved with 4.32\% and 0.3\% errors.

\begin{figure*}[h]
 \centering
\begin{tabular}{c}
  \includegraphics[width=16cm]{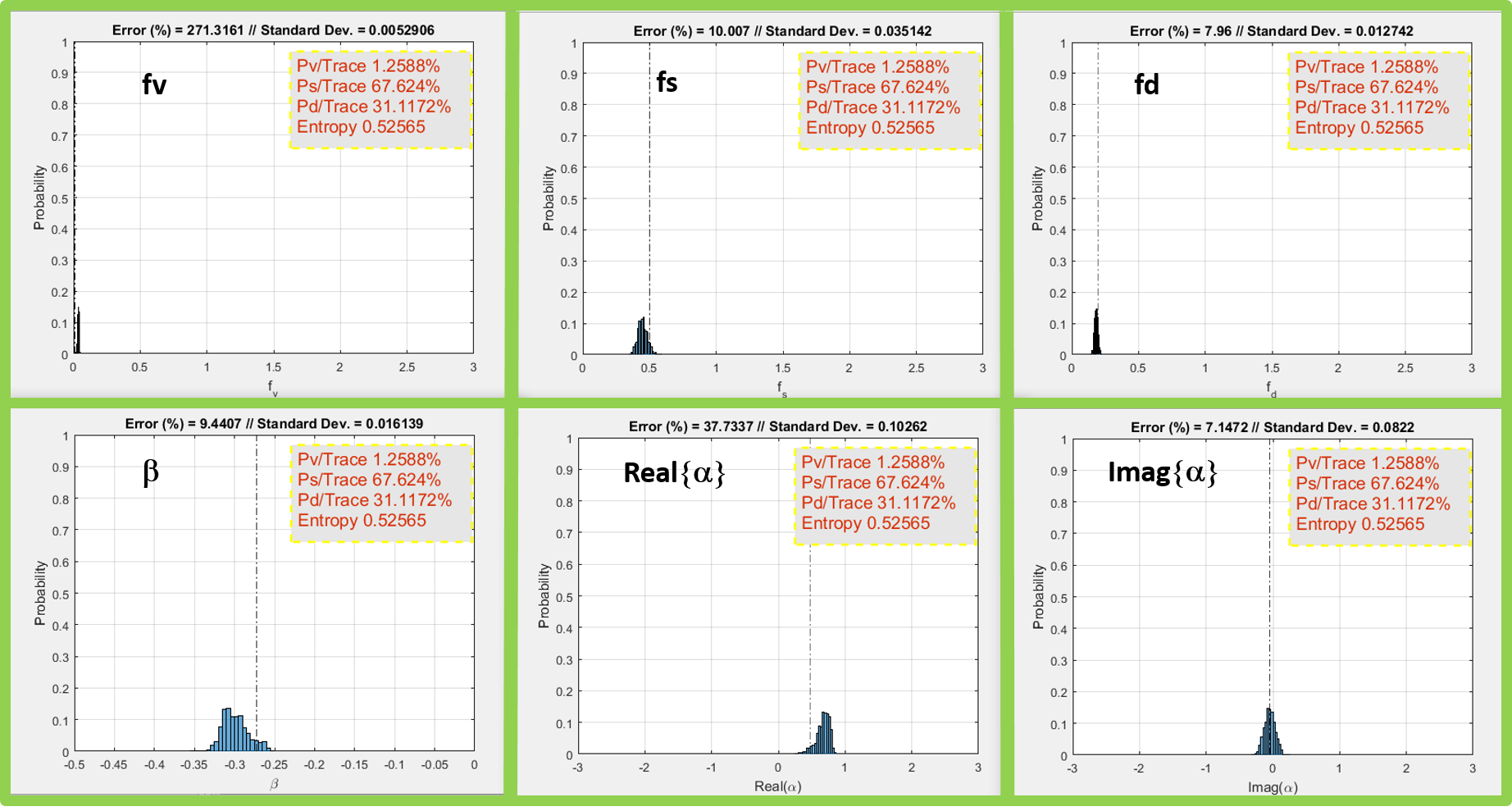}
\end{tabular}
\caption{Low entropy case with higher orientation angle for surface scattering. Boxes inside the plots indicate the relative power for each scattering mechanism and the entropy value. Input parameters: $\epsilon_{soil}=5$, $\theta_0=45^{\circ}$, $\Psi_d=15^{\circ}$ and $\Psi_s=25^{\circ}$}\label{f:low_entropy_2}
\end{figure*}

The second simulation has assumed the same parameters as in the previous case but increasing the orientation angle $\Psi_s$ to 25$^\circ$. As shown in Figure \ref{f:low_entropy_2}, despite being still a low entropy scenario, the estimation error in this case yields high values not only for the the real part of $\alpha$ and $\Psi_s$ but also for $f_s$, $f_d$ and $\beta$ (10\%, 8\%, 9.4\%, respectively). Such inaccuracies lead to relative errors of surface and double-bounce relative backscattering powers of 9\% and 8.62\%, respectively. 

Notwithstanding the inconsistent results obtained in the previous low entropy case, it is pointed out that an increase of the soil effective dielectric constant to a value of 15 (all the other parameter values are maintained) yields an accurate inversion for all parameters and power backscattering values but for the real part of $\alpha$ whose error reaches 18.2\%.

\begin{figure*}
 \centering
\begin{tabular}{c}
  \includegraphics[width=16cm]{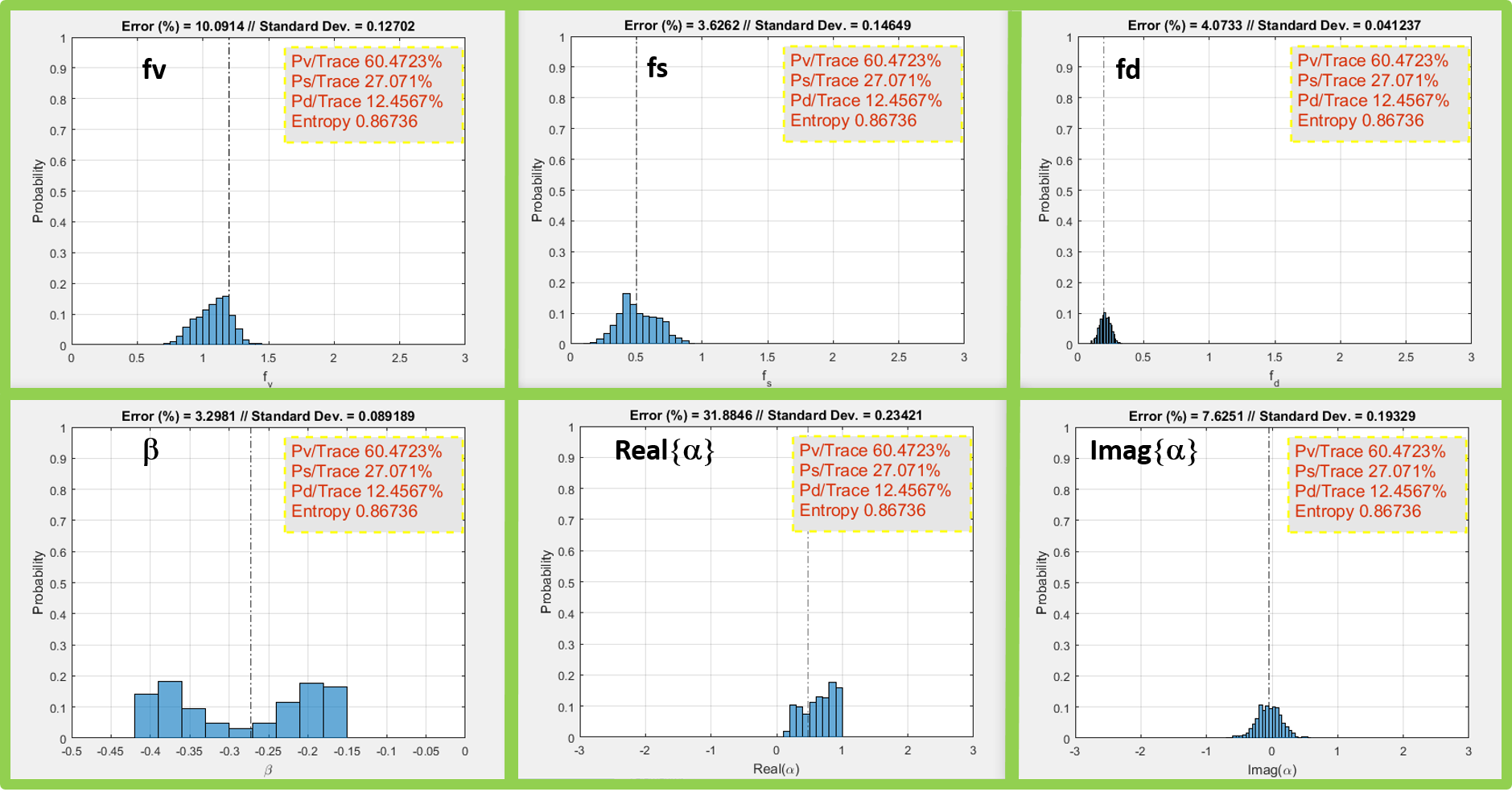}
\end{tabular}
\caption{High entropy case. Boxes inside the plots indicate the relative power for each scattering mechanism and the entropy value. Input parameters: $\epsilon_{soil}=5$, $\theta_0=45^{\circ}$, $\Psi_d=15^{\circ}$ and $\Psi_s=10^{\circ}$}\label{f:high_entropy_1}
\end{figure*}

Next, high entropy scenarios have been assumed. Figure \ref{f:high_entropy_1} illustrates one particular case where the simulated volume, surface and double-bounce relative powers are 60.5\%, 27\% and 12.5\%, respectively. Inconsistencies are found in both the relative error and standard deviations of $f_v$, the real part of $\alpha$ and $\Psi_s$. On the other side, the estimation of the relative backscattering powers for the dominant mechanisms (i.e. volume and surface) yields acceptable results in terms of bias even though their standard deviations exhibit problematic values (see Table \ref{t:high_entropy_1}). 

\begin{table}
\centering
\begin{tabular}{c|c|c} \hline
Rel. Power & Error(\%) & Std.dev. \\ \hline
$P_v/SPAN$ & 10.09 & 6.78\\
$P_s/SPAN$ & 9 & 8.67 \\
$P_d/SPAN$ & 29.1 & 4.77 \\ \hline
\end{tabular}
\caption{High entropy case (Figure \ref{f:high_entropy_1}). Relative error and standard deviation for relative backscattering powers.}
\label{t:high_entropy_1}
\end{table}

\begin{figure*}
 \centering
\begin{tabular}{c}
  \includegraphics[width=16cm]{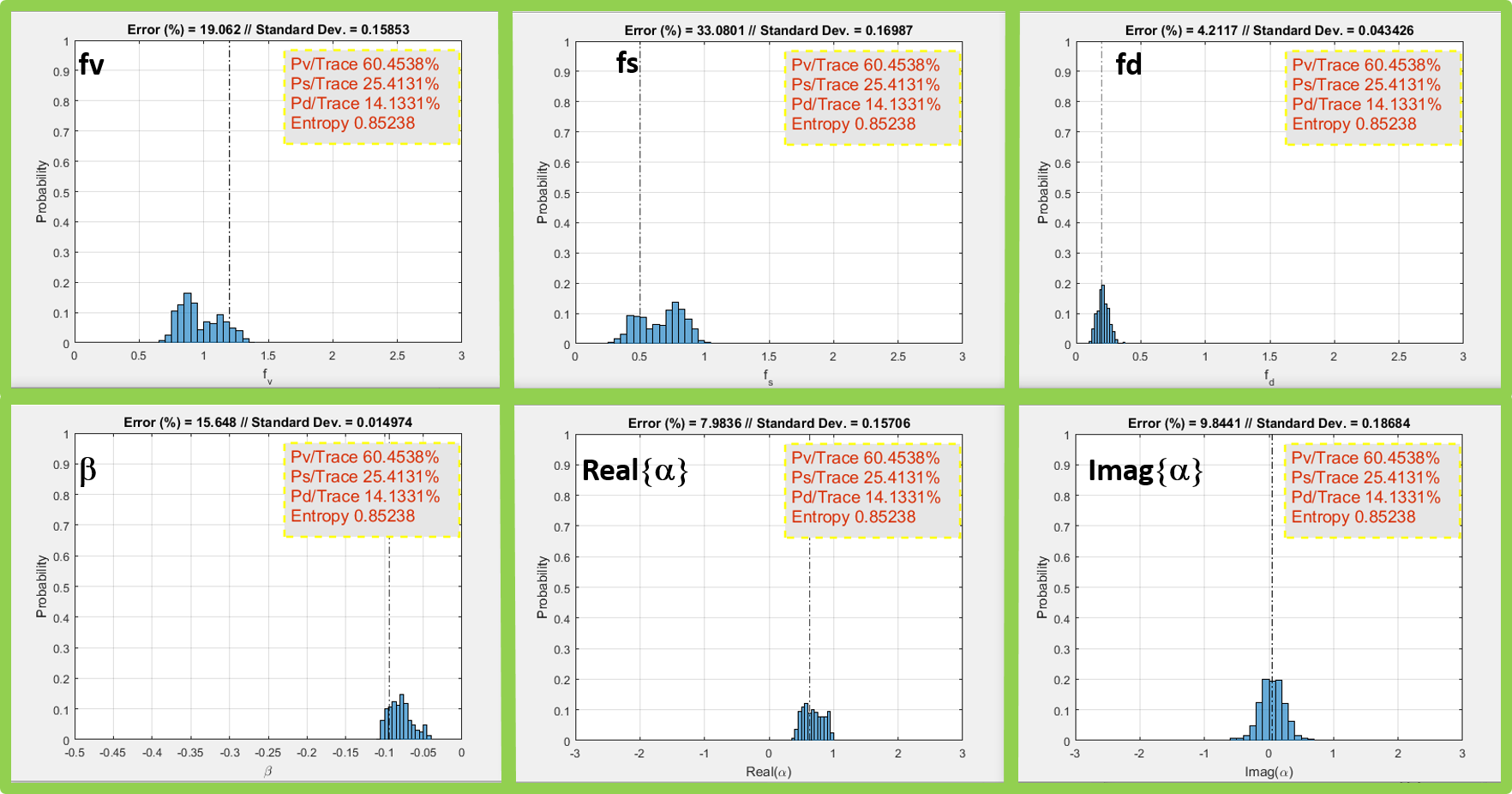}
\end{tabular}
\caption{High entropy case. Box inside the plots indicate the relative power for each scattering mechanism and the entropy value. Input parameters: $\epsilon_{soil}=15$, $\theta_0=21^{\circ}$, $\Psi_d=15^{\circ}$ and $\Psi_s=10^{\circ}$}\label{f:high_entropy_2}
\end{figure*}

\begin{figure*}
 \centering
\begin{tabular}{c}
  \includegraphics[width=13cm]{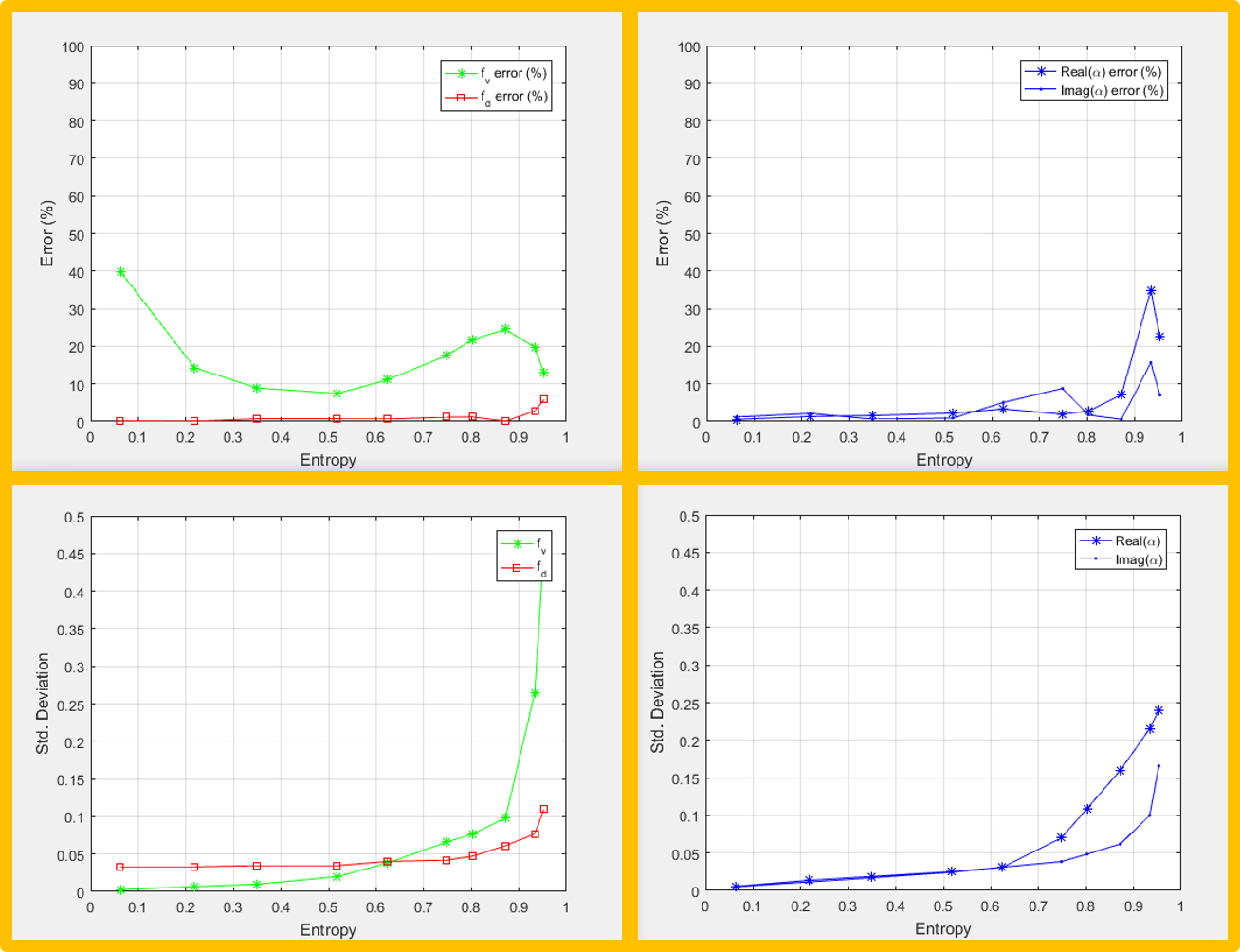}
\end{tabular}
\caption{Estimation performance as a function of entropy considering volume and double-bounce mechanisms. Input parameters: $\epsilon_{soil}=5$, $\theta_0=45^{\circ}$, $\Psi_d=15^{\circ}$ and $\Psi_s=10^{\circ}$}\label{f:vol_dbl}
\end{figure*}

Similarly, we have again simulated a high entropy case (i.e. 0.85) but with a higher soil dielectric constant (i.e 15) and a steeper incidence (i.e. 21$^\circ$). Results are given in Figure \ref{f:high_entropy_2}. Most of parameters are poorly retrieved for this case as well. Such inconsistencies turn into high relative errors for the relative backscattering powers as shown in Table \ref{t:high_entropy_2}.

\begin{table}
\centering
\begin{tabular}{c|c|c} \hline
Rel. Power & Error(\%) & Std.dev. \\ \hline
$P_v/SPAN$ & 17.9 & 7.96\\
$P_s/SPAN$ & 34.5 & 8.54 \\
$P_d/SPAN$ & 14.6 & 3.93 \\ \hline
\end{tabular}
\caption{High entropy case (Figure \ref{f:high_entropy_2}). Relative error and standard deviation for relative backscattering powers.}
\label{t:high_entropy_2}
\end{table}

Finally, we have also considered a different high entropy case where only a dominant volume component and a secondary non-negligible double-bounce mechanism are present. Their relative powers are 75\% and 25\%, respectively. Estimation results show that parameters $f_v$ and the real part of $\alpha$ are not correctly retrieved being their respective errors equal to 13.7\% and 30.6\%. It is noted the huge error in $\alpha$ which in turn would lead to inconsistent estimates for the soil dielectric constant and, consequently, unreliable soil moisture retrievals. 

Table \ref{t:high_entropy_3} provides the error and standard deviation of the relative powers for this case. These results suggest that even the backscattering powers could be compromised according to this simulation as their estimation error is higher than 10\%.

\begin{table}
\centering
\begin{tabular}{c|c|c} \hline
Rel. Power & Error(\%) & Std.dev. \\ \hline
$P_v/SPAN$ & 11.88 & 10.89\\
$P_s/SPAN$ & - & - \\
$P_d/SPAN$ & 10.32 & 4.27 \\ \hline
\end{tabular}
\caption{High entropy case (75\% volume $+$ 25\% double-bounce). Relative error and standard deviation for relative backscattering powers.}
\label{t:high_entropy_3}
\end{table}

The methodology described here can be applied in order to analyse the variation of the estimation accuracy and precision as a function of entropy. As an example, an additional simulation was carried out by considering only volume and double-bounce scattering. The double-bounce scattering was set at a fixed value and the volume coefficient $f_v$ was progressively increased from zero up to a value corresponding to roughly 90\% of the total power. Figure \ref{f:vol_dbl} illustrates the variation of both the relative error and the standard deviation for the two mechanisms considered. Parameters describing the double-bounce mechanism are accurately retrieved up to entropy values around 0.87. On the contrary, the volume coefficient $f_v$ is retrieved with a variable and non-negligible error even for entropy levels down to 0.5-0.6. This same behaviour is translated into the relative backscattering powers.

\section{Discussion}
\label{s:disc}

In order to solve some limitations of the original Freeman-Durden decomposition a number of subsequent works have emerged and different improvements and modifications have been proposed. Usual strategies to assess the performance of such approaches are based on 1) the balance among scattering mechanisms according to theoretical expectations; 2) direct comparison of final products (i.e. soil moisture) with a ground-truth data set collected simultaneously; and 3) interpretation supported on qualitative features extracted by visual inspection (particle orientation and randomness). Despite these approaches have led to consistent outcomes in terms of reliable land cover classification techniques, an in-depth analysis of the whole set of parameters involved in the models has been almost systematically overlooked.

In this work we have pointed out the need to assess the whole set of output parameters to gain a deeper knowledge on the potential limits of applicability of model-based decomposition techniques. The main question we address here regards the physical interpretation of all describing parameters of a particular physically-based PolSAR model. To this aim we have performed a number of simulations according to the general model proposed by Chen et al. \cite{ar:chen2014} which employs all the elements of the coherency matrix. Results from our simulation analysis reveal that not only fine adjustment parameters (i.e. $\beta$, $\alpha$ and orientation angles) but the backscattering powers could be compromised depending upon the scene as estimation errors higher than 10\% for some cases are obtained. We must point out, however, that these conclusions are subject to the particular model employed and the associated inversion approach. In particular, the simulated data used in this study matches the form of the model-based decomposition employed for parameter retrieval. Thus, these results are a \emph{best case scenario} in that the in-scene scatterers exactly match the decomposition model. If the model does not match the in-scene scattering mechanisms additional biases and errors will be introduced. Therefore we acknowledge this study will remain as an anecdotal contribution unless it is replicated by using other general PolSAR models to confirm or not these observations. 

In summary, in the present paper we claim that any new model-based PolSAR approach should be supported by a quantitative assessment on the whole set of parameters provided it was allegedly designed under a \emph{physically-based} basis. To this aim, the use of benchmark datasets available to any researcher and agreed by the community for validation purposes are suggested as a way to better quantify the potential progress by any new proposal.




\bibliographystyle{plain}



%
%
%
\end{document}